\documentclass[sigconf]{acmart}

\usepackage{subfig} 
\usepackage{graphicx} 
\AtBeginDocument{%
  }

\acmYear{2025}\copyrightyear{2025}
\acmConference[MobiSys' 25]{The 23rd Annual International Conference on Mobile Systems, Applications and Services}{June 23--27, 2025}{Anaheim, CA, USA}
\acmBooktitle{The 23rd Annual International Conference on Mobile Systems, Applications and Services (MobiSys' 25), June 23--27, 2025, Anaheim, CA, USA}
\acmDOI{10.1145/3711875.3736681}
\acmISBN{979-8-4007-1453-5/25/06}

\begin{document}

\title{Privacy-Aware Ambient Audio Sensing for Healthy Indoor Spaces}


\author{Bhawana Chhaglani}
\affiliation{%
  \institution{University of Massachusetts Amherst}
  \city{}
  \country{}
  }
\email{bchhaglani@umass.edu}



\renewcommand{\shortauthors}{Chhaglani et al.}

\begin{abstract}
\normalsize

Indoor airborne transmission poses a significant health risk, yet current monitoring solutions are invasive, costly, or fail to address it directly. My research explores the untapped potential of ambient audio sensing to estimate key transmission risk factors such as ventilation, aerosol emissions, and occupant distribution—non-invasively and in real time. I develop privacy-preserving systems that leverage existing microphones to monitor the whole spectrum of indoor air quality which can have a significant effect on an individual's health. This work lays the foundation for privacy-aware airborne risk monitoring using everyday devices.

\end{abstract}






\maketitle

\vspace{-0.25cm}
\section{Introduction}

\begin{figure}[t]
    \centering
    \includegraphics[width=\linewidth]{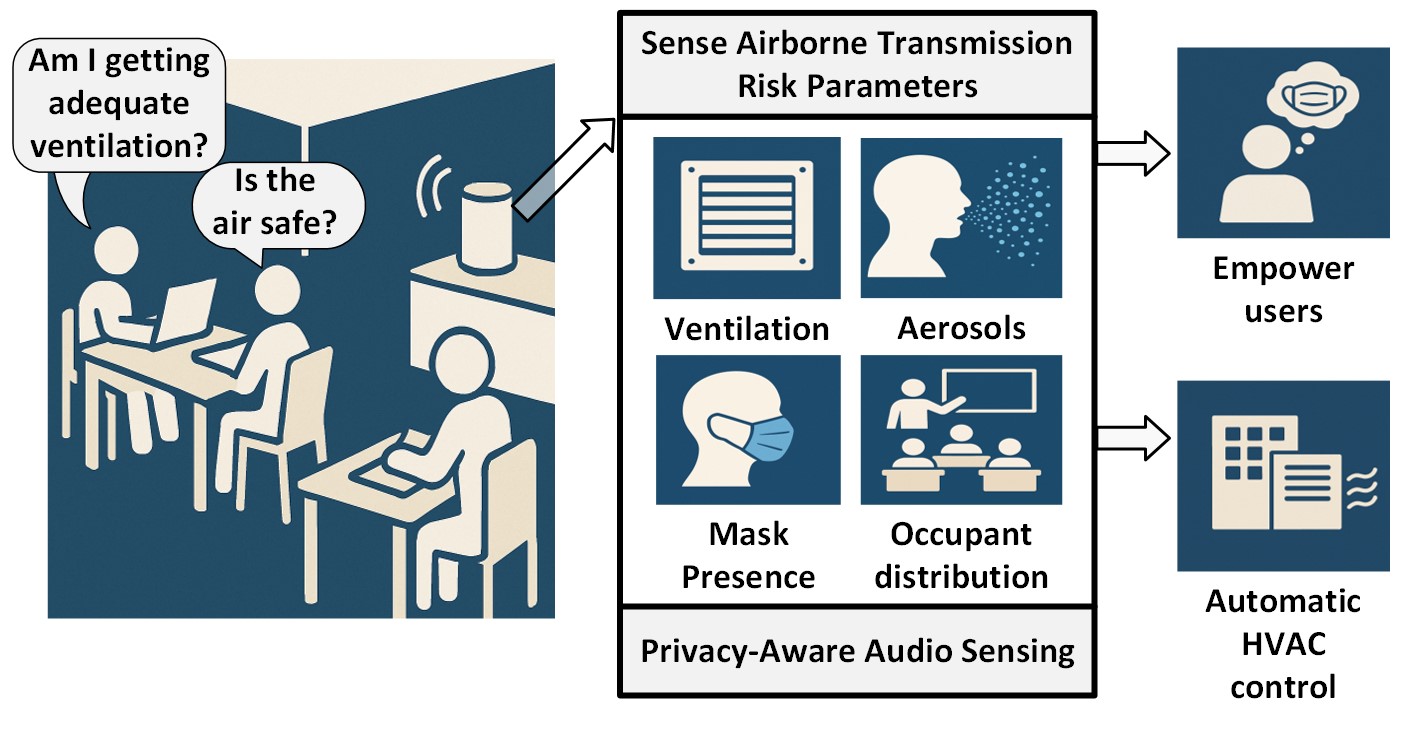}
    \caption{Overview of privacy-aware audio sensing for indoor airborne risk assessment. }
    \label{fig:concept}
    \vspace{-0.3cm}
\end{figure}

Have you ever wondered if the air around you is truly safe? Or chosen where to eat based on how well-ventilated a space is? We spend over 90\% of our time indoors, making indoor air quality critically important—especially during flu season, a pandemic, or for immunocompromised individuals. Airborne transmission of infectious diseases occurs when viral or bacterial particles linger in the air and are inhaled by others. This risk increases indoors, where viral particles are more likely to accumulate due to limited airflow and confined spaces.
While organizations like WHO, CDC, and ASHRAE \cite{ashraeOpen} emphasize the importance of proper ventilation, existing research in smart buildings primarily focus on optimizing energy efficiency and thermal comfort \cite{lam2014occupant}, not minimizing the risk of airborne disease transmission. Current indoor air quality efforts tend to target pollutants like dust, mold, or VOCs, leaving a significant gap in monitoring and mitigating infectious airborne agents. Simply increasing ventilation, as was done during the pandemic, is not sustainable long-term due to its high energy cost.
This creates a clear airborne risk mitigation gap. Having learned from the pandemic, we now need systems that can estimate and communicate the real-time risk of indoor airborne transmission—helping people ask and answer critical questions like “Is this space well-ventilated?” or “Is the air likely virus-free?” This will empower users with critical indoor air information and encourage building management to maintain standards.
\\Sensing airborne transmission risk is inherently challenging as it depends on a complex interplay of factors such as ventilation, aerosol emissions, mask usage, and occupant distribution. Existing solutions to monitor these parameters are often invasive, expensive, or impractical for everyday use \cite{ee650_air_flow_sensor,CPC3775}. My research addresses this gap by demonstrating how audio sensing offers a unique, non-invasive opportunity to monitor key transmission risk indicators (like ventilation and aerosols) using ubiquitous microphones \cite{chhaglani2023cocoon}. 
However, audio presents significant privacy concerns, especially if we want to deploy microphone-based system in indoor spaces like offices and conference rooms. To overcome this, I design privacy-preserving sensing systems that extract relevant environmental parameters without accessing raw audio or compromising user privacy. 
To this end, I make the following contributions.
My work introduces \textit{FlowSense} \cite{chhaglani2022flowsense}, a system that monitors ventilation using low-frequency vent noise; \textit{AeroSense} \cite{chhaglani2024aerosense,chhaglani2023breatheasy}, which predicts aerosols from human-respiratory activities (like speaking or coughing) and predicts airborne risk \cite{chhaglani2023breatheasy}; and \textit{FeatureSense} \cite{chhaglani2024towards}, a general-purpose privacy-aware feature library that assess and prevents speech and speaker leakage. 
While designing these systems, I address the challenges of privacy, ambient noise interference, and robustness to user and environment factors. 
Together, these systems illustrate the potential of ambient audio sensing to enable scalable, real-time, and privacy-aware airborne risk monitoring.

\vspace{-0.2cm}
\section{\Large Challenges and Contributions}
\textbf{Ventilation Sensing}: Ventilation is crucial, but less than 20\% of commercial buildings in the United States use Building Automation System (BAS). 
Moreover, it is very expensive to retrofit the HVAC systems with these sensors and still the information is not accessible to occupants. 
The other techniques to sense ventilation are pressure sensors or vane anemometer \cite{ee650_air_flow_sensor}, but this requires removing vents and placing sensors inside duct, which is not feasible.
\textit{FlowSense} \cite{chhaglani2022flowsense} is a low-cost, easy to deploy, non-intrusive solution that estimates ventilation rate by listening to low-frequency airflow sounds from HVAC vents. It employs silent period detection and optimal low pass filtering to ensure user privacy.  Additionally, we propose the Minimum Persistent Sensing (MPS) as a post-processing algorithm to reduce interference from ambient noise (e.g., office machines).
The system achieves over 99\% accuracy in predicting vent status and 0.5 MSE in predicting airflow rate. FlowSense transforms smartphones and ambient microphones into airflow sensors, empowering users and building management to assess ventilation adequacy in real time.
\\ \textbf{Aerosol and Airborne Risk Sensing}: Aerosols are key indicator of airborne transmission because they remain suspended in air, travel long distances, and carry infectious agents. However, detecting aerosols is extremely challenging as the aerosol sensor \cite{CPC3775} is bulky, expensive, and designed for laboratory use. 
Since aerosol emissions depend on the type and level of respiratory activity \cite{asadi2019aerosol}, we design \textit{AeroSense} \cite{chhaglani2024aerosense} that detects human activities (e.g., speaking, coughing \cite{chhaglani2024cough}, sneezing) using ambient acoustic sensing. 
Using non-reconstructible audio features captured from microphone arrays, AeroSense detects and localizes human respiratory activities that contribute to aerosol emissions. It estimates aerosol emission in real-time with 2.34 MSE and 0.73 R². Since mask reduces aerosol emissions by 70\%, we use feature engineering to detect the presence of the mask. We also design features to detect voice liveness \cite{chhaglani2025livedetector} as the aerosols emitted from electronic speaker (like zoom calls) should not be counted. 
Additionally, we use signal processing techniques to detect interpersonal distances, which is another important parameter of airborne risk assessment. 
We demonstrate efficacy of AeroSense through controlled and in-the-wild experiments. 
BreathEasy \cite{chhaglani2023breatheasy} serves as a conceptual framework that unifies aerosol and ventilation sensing capabilities into a holistic airborne risk sensing pipeline. It demonstrates the feasibility of sensing multiple parameters—ventilation, activity type and loudness, occupant distribution, and mask usage—using only audio, and use it to assess risk of transmission. Collectively, these systems form a novel class of non-intrusive and practical solutions to assess indoor airborne transmission risk in real time and optimize ventilation strategies.

\noindent \textbf{Privacy-Aware Audio Sensing}:
Despite its potential, audio sensing raises significant privacy concerns. Most existing privacy-aware methods focus solely on obfuscating or suppressing speech \cite{boovaraghavan2024kirigami}, however audio contains more privacy sensitive information than just speech, such as speaker identity and whereabouts as described in \cite{chhaglani2024towards}.
To address broader privacy concerns in audio sensing, I propose \textit{FeatureSense} \cite{chhaglani2025featuresense} library which exposes generalizable privacy-aware audio features that reduce speech and speaker-related information leakage while maintaining the effectiveness of audio-based applications. By designing a comprehensive privacy evaluation framework and adaptive task-specific feature selection, this work provides a foundational framework for ensuring trust in audio sensing technologies \cite{chhaglani2024towards}. Across multiple tasks (e.g., cough detection, environmental sound classification), FeatureSense outperforms existing techniques, achieving up to 60.6\% reduction in speaker leakage without sacrificing utility.
This line of work creates a foundational privacy framework for future audio-based sensing systems, ensuring trust and accountability in ambient intelligence applications.

\vspace{-12.6pt}
\section{The Road Ahead}
My work advances non-invasive, privacy-preserving audio sensing for health and safety.
I develop interpretable models that link specific acoustic signatures to airflow noise, aerosol emissions, or privacy-sensitive attributes. Unlike black-box AI, this approach enhances transparency, adaptability, and makes sensing more robust. 
This research represents an important step towards explainable and deployable audio sensing systems that unlock the potential of this rich data modality while ensuring user privacy.\\
My research will further explore the intersection of audio sensing, privacy, and public health. One promising direction is to detect the directionality of airflow using acoustic signals, which would enable identification of poorly ventilated zones. I also plan to develop techniques for HVAC system fault diagnostics by learning acoustic signatures of healthy versus faulty systems. 
Another exciting avenue is to explore the causality between audio and aerosol emissions, enabling more precise estimation of aerosol emissions based on acoustic features of human activity \cite{asadi2019aerosol}. These efforts will deepen scientific understanding of aerosol transmission and support robust, explainable sensing models.
Looking forward, I envision building a robust ecosystem of ambient intelligence that integrates audio with other low-cost modalities to provide holistic environmental awareness. Currently, my research explores two main directions:\\
\noindent \textbf{Predicting crowd densities using passive WiFi logs}: A natural extension of my work is to monitor and predict crowd densities in indoor spaces by integrating ambient audio with passive WiFi logs. While audio provides insights into activity and respiratory emissions, WiFi logs can passively capture spatial and temporal occupancy trends. By fusing these two modalities, we can gain a richer picture of crowd presence and crowd type (e.g., static vs dynamic), which is critical for assessing airborne risk. This multimodal approach enables the detection of high-risk zones and supports context-aware decision-making—encouraging users to wear masks, ventilate spaces, or avoid certain areas altogether.
Building on this idea, I am developing \textit{CrowdSense}, a micro-foundational model that leverages passive WiFi logs to predict crowd densities and provide personalized alerts based on users’ spatial trajectories. 


\noindent \textbf{Evaluating the Privacy of Feature-Based Audio Sensing}: While feature-based approaches offer a promising route for privacy-aware audio sensing, it is hard to quantify its privacy leakage. I am investigating the extent of phoneme-level leakage from these handcrafted "safe" features by attempting to reconstruct audio to infer phoneme leakage, which can then be passed through LLMs to infer complete utterances. This line of work not only questions current assumptions about the safety of low-level features but also emphasizes the need for a comprehensive privacy evaluation framework.
This work will help establish standardized benchmarks for evaluating privacy leakage in sensing systems and inform future regulations around audio sensing in public and private spaces \cite{chhaglani2025artifree}.\\
Together, these future directions will enable ambient intelligence systems that are not only effective and scalable but also ethical and respectful of user privacy—ushering in a new era of privacy-aware, health-driven sensing for smart environments \cite{chhaglani2024neckcare,chhaglani2025neckcheck, chhaglani2025vision, chhaglani2025privacy,aziz2024hardware,iyengar2021holistic,agrawal2024oas}.

\vspace{-0.5cm}
\section{Biography}
Bhawana Chhaglani is a fifth-year Ph.D. candidate at the University of Massachusetts Amherst. She is advised by Prof. Prashant Shenoy and Prof. Jeremy Gummeson. 

\footnotesize
\bibliographystyle{ACM-Reference-Format}
\bibliography{main}

\end{document}